\documentclass[pre,reprint,amsmath,amssymb,showpacs]{revtex4-1}
\usepackage{dcolumn}
\usepackage{bm}
\usepackage[dvips]{color}
\usepackage{subfigure}
\usepackage{graphicx} 
\usepackage{algorithmicx}
\usepackage[ruled]{algorithm}
\usepackage{algpseudocode}
\newcommand{\rvec}{\textit{\textbf{r}} }
\newcommand{\vvec}{\textit{\textbf{v}} }
\newcommand{\evec}{\textit{\textbf{e}} }
\newcommand{\qvec}{\textit{\textbf{q}} }
\newcommand{\ivec}{\textit{\textbf{i}} }

\newcommand{\Xvec}{\textit{\textbf{X}} }
\newcommand{\fvec}{\textit{\textbf{f}} }
\newcommand{\Fvec}{\textit{\textbf{F}} }
\newcommand{\Evec}{\textit{\textbf{E}} }
\newcommand{\Rmat}{\textsf{\textbf{R}} }
\newcommand{\Kmat}{\textsf{\textbf{K}} }

\newcommand{\Imat}{\textsf{\textbf{I}} }
\newcommand{\Gmat}{\textsf{\textbf{G}} }
\newcommand{\Cmat}{\textsf{\textbf{C}} }
\newcommand{\Tmat}{\textsf{\textbf{T}} }

\begin{document}

\title{Versatile low-Reynolds-number swimmer with three-dimensional maneuverability}

\author{Mir Abbas Jalali$^{1,2}$}
\email{mjalali@berkeley.edu}
\author{Mohammad-Reza Alam$^{3}$}
\email{reza.alam@berkeley.edu}
\author{SeyyedHossein Mousavi$^2$}

\affiliation{
$^1$Department of Astronomy, University of California, Berkeley, CA 94720, USA \\
$^2$Department of Mechanical Engineering, Sharif University of Technology, Azadi Avenue, 
Tehran, Iran \\
$^3$Department of Mechanical Engineering, University of California, Berkeley, California 94720, USA}

\date{\today}

\begin{abstract}
We design and simulate the motion of a new swimmer, the {\it Quadroar}, with three 
dimensional translation and reorientation capabilities in low Reynolds number conditions. 
The Quadroar is composed of an $\texttt{I}$-shaped frame whose body link is a simple linear  
actuator, and four disks that can rotate about the axes of flange links. The time symmetry 
is broken by a combination of disk rotations and the one-dimensional expansion/contraction 
of the body link. The Quadroar propels on forward and transverse straight lines and 
performs full three dimensional reorientation maneuvers, which enable it to swim along 
arbitrary trajectories. We find continuous operation modes that propel the swimmer on 
planar and three dimensional periodic and quasi-periodic orbits. Precessing quasi-periodic
orbits consist of slow lingering phases with cardioid or multiloop turns followed by directional 
propulsive phases. Quasi-periodic orbits allow the swimmer to access large parts of its neighboring 
space without using complex control strategies. We also discuss the feasibility of fabricating 
a nano-scale Quadroar by photoactive molecular rotors.
 
\end{abstract}


\maketitle

\section{Introduction}

The dynamics of a self-propelling object in a viscous fluid is characterized 
by the Reynolds number that determines the relative importance of inertia forces against 
viscous ones. When the Reynolds number is small, the exchange of momentum between 
the object and the background fluid is small and the object is driven mainly by drag forces. 
The frictional interaction between the object and the background fluid is only a necessary 
condition for swimming at low Reynolds numbers, for the Stokes equation governing the 
flow of the background fluid remains invariant under the reflection transformations 
$t\rightarrow -t$ over the time domain, and consequently, a swimmer cannot advance 
in the configuration space through reciprocal movements or deformations of its body. 
This is why swimmers consisting of a finite number of rigid components need at least 
two degrees of freedom to break the time symmetry and stroke \cite{purcell,Lauga}. 

Artificial low-Reynolds-number swimmers have applications in both the micro 
and macro scales. In micron-scales, they can deliver drug and cargo to cells, and 
interact with micro-organisms and bacteria. In large macro-scales they can conduct inspection 
missions in oil tanks and highly viscous reservoirs. Like any robot, and regardless of the 
mechanism and swimming strategy, the major technological challenge of building 
swimmers in low Reynolds number conditions is how to power their actuators and 
control them onboard or remotely. Therefore, not all proposed swimmers can be 
engineered and operated in reality. The simplest low-Reynolds-number swimmer is 
the linked three-sphere of \citet{NG04} that rectilinearly swims using two telescopic 
actuators. The hydrodynamics, global controllability and optimal stroking of the 
Najafi--Golestanian swimmer have been studied in detail \cite{GA08,APY09,ADL08} 
and experiments with an optical tweezer have shown that the three-sphere system 
can indeed generate a net flow \cite{L09}, necessary for swimming. The rotating 
version of this swimmer has been proposed in \cite{DBS05}. Pushme-pullyou is 
another promising rectilinear swimmer \cite{AKO05} that uses one telescopic 
actuator and two spherical bladders that exchange their volume in each stroke. 

Low-Reynolds-number swimming in two or three dimensions can be performed 
using triangular or tetrahedral configurations of linked spheres \cite{LLY12,A11} 
and two-sphere propellers \cite{NZ10}. The main advantage of the circle swimmer 
proposed in \cite{LLY12} is that it can move on complex planar trajectories only 
by controlling the angle between its two links. Deformable bodies \cite{SW89,AGK04,SBH12} 
and articulated robots with rigid links \cite{ADGZ13} have also been suggested for 
low-Reynolds-number swimming. Nonetheless, the realization of swimmers with 
deformable surfaces and with intuitive bladders as in pushme-pullyou \cite{AKO05} 
is a technological challenge, especially for two- and three-dimensional swims. 
Moreover, swimmers with many linked spheres and articulated $N$-link tails are 
not optimum systems because of two reasons: they require a complex control strategy 
and too many degrees of freedom to perform a planned maneuver, and they experience 
lateral drifts in each stroke by generating unnecessary flows perpendicular to the 
motion direction. 

In this study, we propose the concept and analyze the motion of a swimmer with five 
degrees of freedom that combines the rotations of four disks (oars) and the telescopic 
motion of its body link, so follows the name {\it Quadroar}\,=Quadru+oar, to perform full 
three dimensional movements. The swimmer has three main features: (i) It can perform 
reorientation maneuvers by rotating about all of its three principal body axes, and stroke 
along prescribed three dimensional curves without unnecessary side drifts. 
(ii) Its concept can be realized from nano to macro scales, with 
a variety of scientifically feasible and technologically available actuation mechanisms. 
(iii) Similar to what the circle swimmer \cite{LLY12} can perform in two dimensions, 
it can track prescribed 3D paths using stepwise control strategies, or move on 
quasi-periodic orbits by the continuous operation of its actuators. Quasi-periodic 
orbits become dense in certain regions of the configuration space, and enable the 
swimmer to access large parts of its neighboring space. 

The paper is organized as follows. We introduce the geometry and degrees of freedom 
of the Quadroar in section \ref{sec:geometry} and derive its equations of motion in 
section \ref{actuation modes} where we also discuss its independent operation 
modes that lead to directional motions and rigid body maneuvers by pulsed commands
to its actuators. Section \ref{sec:frequency-tuning} is dedicated to swimming modes 
with continuous operations of main actuators. We also explore the parameter space 
to generate a wide range of quasi-periodic orbits in two and three dimensions. 
Section \ref{sec:discuss} concludes the paper by discussing the unique features of 
the Quadroar, its possible applications, and its realization in nano scales.

\section{Geometry and kinematics}
\label{sec:geometry}

\begin{figure}
\centerline{\hbox{\includegraphics[width=0.23\textwidth]{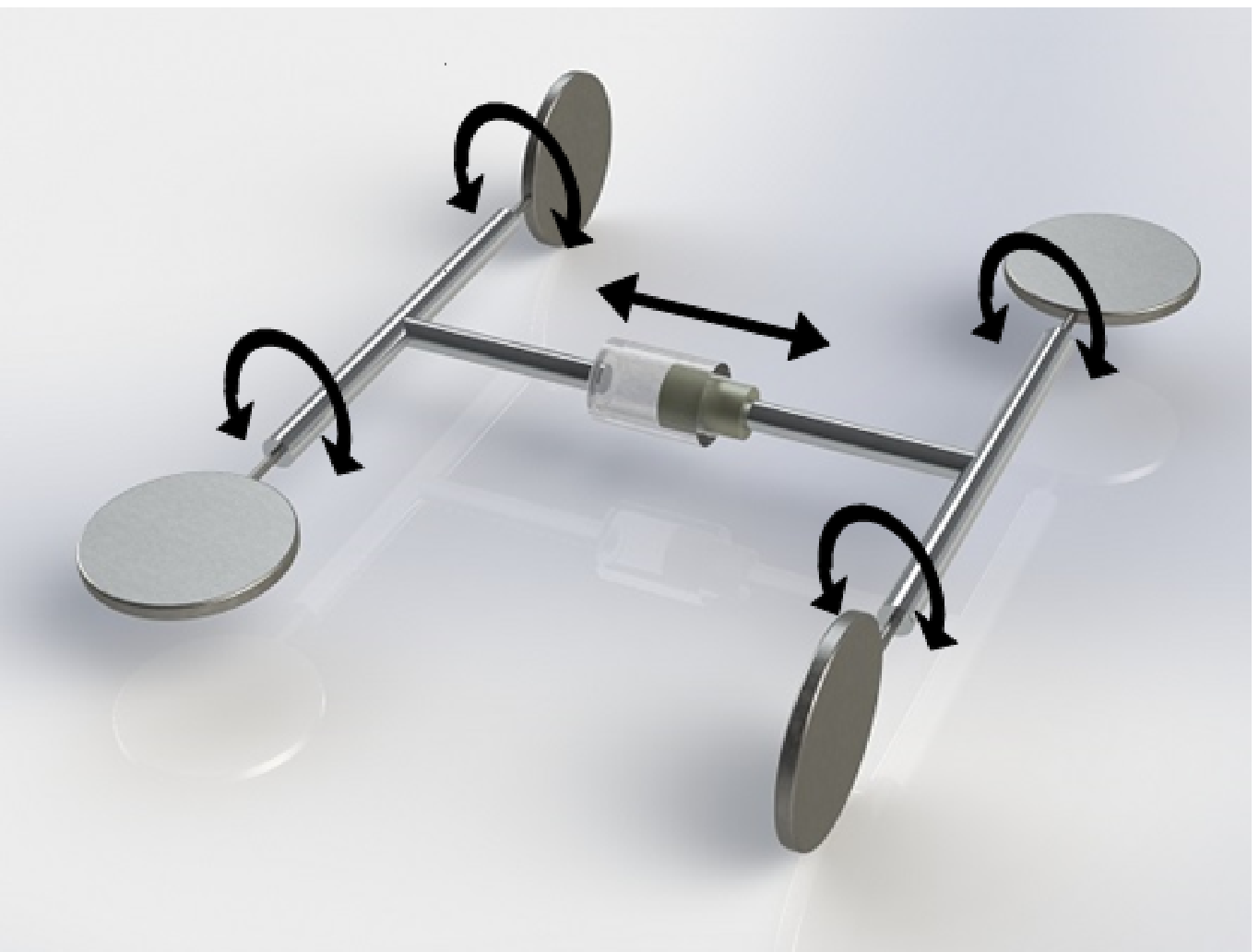} }
\hbox{\includegraphics[width=0.22\textwidth]{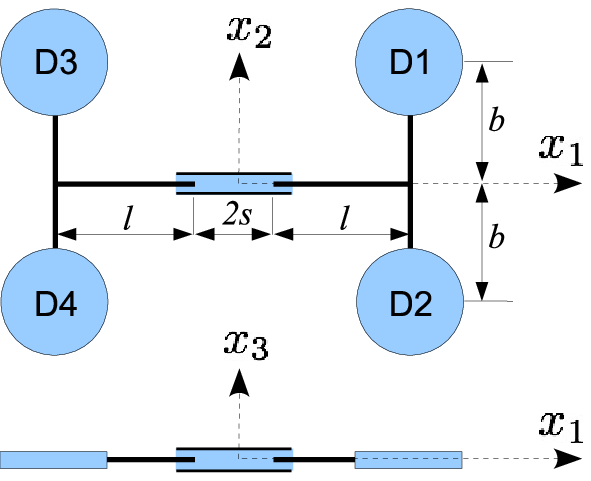} }
 }
\caption{The geometry of the Quadroar. The body link of the $\texttt{I}$-frame 
is a linear actuator, here schematically shown as a telescopic component. Four disks 
are mounted at the end points of flange links. Each disk has a single rotational degree 
of freedom about the axis of its supporting link. The body-fixed coordinate frame is such 
that the \texttt{I}-frame lies in the $(x_1,x_2)$-plane.}
\label{fig1}
\end{figure}

Figure \ref{fig1} shows the geometry of Quadroar that consists of three links with 
an $\texttt{I}$-shape arrangement and four similar disks of radius $a$. The length 
of the body link continuously varies by a linear actuator, and disks can rotate 
continuously or stepwise about the axes of flange links. From here on, we refer to 
the set of the links as the ``\texttt{I}-frame". The length of the body link is $2l+2s$ 
where $2s$ is the contribution from the expansion/contraction of the linear actuator. 
The front and rear flange links have constant lengths of $2(b-a)$ (the distance from 
the cross section of the body link and a flange link to the center of each disk is $b$). 
The rotation of the $n$-th disk ($n=1,2,3,4$) about its supporting flange link is 
measured by the angle $-\pi \le \vartheta_n \le \pi$ that the plane of the disk 
makes with the $\texttt{I}$-frame.

We denote the inertial coordinate system and its corresponding basis vectors by 
$(X_1,X_2,X_3)$ and $(\Evec_1,\Evec_2,\Evec_3)$, respectively. 
The translational motion of the swimmer is thus described by the position of its 
center of mass, $\Xvec_{\rm c}=(X_{1{\rm c}},X_{2{\rm c}},X_{3{\rm c}})$, 
and its orientation is determined by the roll-pitch-yaw angles 
$\boldsymbol{\alpha}=(\phi,\theta,\psi)$. Consider a body-fixed coordinate frame 
$(x_1,x_2,x_3)$, with the unit vectors $(\evec_1,\evec_2,\evec_3)$, whose $x_1$-axis 
is along the body link and $x_3$-axis is perpendicular to the $\texttt{I}$-frame
(Fig. \ref{fig1}). Defining $\Rmat_{x_i}(\gamma)$ as a rotation matrix about the $x_i$-axis 
by the angle $\gamma$, transformation from $\Evec_i$ to $\evec_i$ ($i=1,2,3$) is carried 
out using the successive operation of the matrices $\Rmat_{x_3}(\psi)$, 
$\Rmat_{x_2}(\theta)$ and $\Rmat_{x_1}(\phi)$ (see Appendix \ref{app:A}), 
and we obtain
\begin{eqnarray}
\Evec_i = \Rmat \cdot \evec_i, ~~
\Rmat \equiv R_{ij} \evec_i \evec_j = \Rmat_{x_1} \cdot \Rmat_{x_2} \cdot \Rmat_{x_3}.
\end{eqnarray}
Here and throughout the paper, repeated indices associated with the components of 
vectors or elements of tensors denote summations over those indices. Assuming that the 
linear actuation of the body link does not change the location of swimmer's center 
of mass, and denoting the angular velocity of the swimmer in the body frame by 
$\boldsymbol{\Omega}$, the velocity of the $n$th disk measured at its center of 
mass is obtained from  
\begin{eqnarray}
\vvec_n &=& \Rmat \cdot \vvec_{\rm c} + 
\dot \ell_n \evec_1+\boldsymbol{\Omega} \times \rvec_n,~~\vvec_{\rm c}=\dot \Xvec_{\rm c}, 
\label{eq:vs} \\
\rvec_n &=& r_{n,i} \evec_i = \ell_n \evec_1 + b_n \evec_2,
\end{eqnarray}
where the over-dot denotes $d/dt$ and
\begin{eqnarray}
\ell_n &=& \left \{
\begin{array}{cc}
+ (l+s) ~~{\rm for}  &  n=1,2 \\
- (l+s) ~~{\rm for}   & n=3,4
\end{array}
\right. , \\
b_n &=& \left \{
\begin{array}{cc}
+b~~{\rm for}  &  n=1,3  \\
-b~~{\rm for}   & n=2,4 
\end{array}
\right. ; 
\end{eqnarray}
The linear actuator expands when $\dot s>0$ and contracts when $\dot s<0$. 
For the sake of compact notation, let $s_\gamma$ and $c_\gamma$ denote 
$\sin(\gamma)$ and $\cos(\gamma)$, respectively. Since the disks can rotate 
with respect to the frame, their absolute angular velocities become
\begin{eqnarray}
\boldsymbol{\omega}_n = \boldsymbol{\Omega}+\dot \vartheta_n \evec_2,
\label{eq:omegas}
\end{eqnarray}
where 
\begin{eqnarray}
\boldsymbol{\Omega} = \Tmat \cdot \dot{\boldsymbol{\alpha}}, ~~
\Tmat = 
\left [
\begin{array}{ccc}
1  & 0  & -s_\theta   \\
0  & c_\phi  & c_\theta s_\phi   \\
0  & -s_\phi  & c_\theta c_\phi   
\end{array}
\right ].
\label{eq:omega-vs-angleDots}
\end{eqnarray}
The transformation matrix $\Tmat$ has a coordinate type 
singularity (gimbal lock) at $\theta=\pm \pi/2$. This singularity is avoided by carrying
out attitude computations in the space of unit quaternions. Dynamics of quaternions 
and their relationship with roll-pitch-yaw angles are discussed in Appendix \ref{app:B}. 
With five control inputs, $s(t)$ and $\vartheta_n(t)$, the Quadroar is an under-actuated 
system that can move along, and rotate about, any of its body axes except performing 
side slip along the $x_2$ axis. In section \ref{actuation modes} we will show that being 
under-actuated does not restrict the three-dimensional maneuvering and tracking 
capability of the Quadroar.

\section{Swimming and reorientation}
\label{actuation modes}

We assume that the effective cross section (area) of the \texttt{I}-frame is negligible 
compared to four disks, and therefore, the drag forces and torques that the frame 
links experience are ignored. The needed driving force for translation and the 
torque for rigid-body rotation come from the disks, which feel fluid resistance as 
they move and rotate. We also work with a mechanism that preserves the location 
of the swimmer's center of mass during the actuation of the body link. One more 
simplifying assumption is that the disks are apart enough so that the fluid 
streaming around each disk is not affected by the movement of the frame 
or other disks. This requirement confines us to models with $a \ll l,b$.

For a background fluid of dynamic viscosity $\mu$, and in the absence of accelerations, 
the translation tensor of a circular disk is given by equation (5-4.29) in \cite{HB}:
\begin{eqnarray}
\Kmat = \left ( \frac{32}{3} \ivec_1\ivec_1 + \frac{32}{3} \ivec_2\ivec_2 +16 \ivec_3\ivec_3  \right ) a,
\label{eq:K}
\end{eqnarray}
where $(\ivec_1,\ivec_2,\ivec_3)$ define an orthogonal coordinate system 
fixed to the disk. The origin of the coordinate system coincides with the center 
of mass of the disk and $\ivec_3$ is normal to the disk plane. The drag force 
that the $n$th disk experiences as it moves with the velocity $\vvec_n$ thus 
reads
\begin{eqnarray}
\Fvec_n = -\mu \Kmat \cdot \vvec_n.
\end{eqnarray}
When the $n$th disk is actuated, its axis of rotation always remains along the 
$x_2$-axis of the \texttt{I}-frame. It is therefore a computational convenience to
set $\ivec_2=\evec_2$ and obtain
\begin{eqnarray}
\left (
\begin{array}{c}
\ivec_1  \\
\ivec_2  \\
\ivec_3   
\end{array}
\right ) =
\left [
\begin{array}{ccc}
\cos (\vartheta_n) ~ &  0  & ~ \sin (\vartheta_n)  \\
0 ~ & 1   & ~ 0  \\
-\sin (\vartheta_n) ~ & 0  & ~ \cos (\vartheta_n)  
\end{array}
\right ] \cdot 
\left (
\begin{array}{c}
\evec_1  \\
\evec_2  \\
\evec_3   
\end{array}
\right ).
\label{eq:i-vs-e}
\end{eqnarray}
We note that the rotation angle $\vartheta_n$ is positive if its angular velocity vector 
is along the positive $x_2$-axis. Substitution of (\ref{eq:i-vs-e}) into (\ref{eq:K}) results 
in the translation tensor $\Kmat_n = K_{n,ij} \evec_i \evec_j$ of the $n$th disk with 
respect to the frame of the swimmer:
\begin{eqnarray}
\Kmat_n = \frac 83 a \left [
\begin{array}{ccc}
5-\cos \left (2 \vartheta_n \right ) ~ & 0 & ~ \sin \left ( 2 \vartheta_n \right ) \\
0  & 4  & 0  \\
\sin \left ( 2\vartheta_n \right ) ~  &  0  & ~ 5 + \cos \left ( 2\vartheta_n \right )
\end{array}
\right ].
\end{eqnarray}
In the low-Reynolds regime the acceleration of the swimmer is small. Therefore, 
if the swimmer is designed to be neutrally buoyant, the sum of the drag forces 
$\Fvec_n=-\mu \Kmat_n\cdot \vvec_n$ vanishes and we obtain the force balance 
equation
\begin{eqnarray}
\mu \sum_{n=1}^{4} \Kmat_n \cdot \vvec_n = 0. \label{eq:force-balance}
\end{eqnarray}

The resistive torque that a circular disk experiences due to its pure rotation is calculated 
from \cite{HB}
\begin{eqnarray}
\boldsymbol{\tau}_n=- \mu \Gmat \cdot \boldsymbol{\omega_n}, ~~\Gmat=\frac {32}{3} a^3 \Imat,
\end{eqnarray}
where $\Imat$ is the identity matrix and $\Gmat$ is the rotation tensor, which is isotropic at the 
center of the disk. The resultant torque exerted on the swimmer is the superposition of $\boldsymbol{\tau}_n$ 
with the torques $\rvec_n\times \Fvec_n$ that the drag forces generate about the center of mass of 
the swimmer. Consequently, the torque balance equation becomes 
\begin{eqnarray}
\mu \sum_{n=1}^{4} \left [ \Gmat \cdot \boldsymbol{\omega}_n 
+ \rvec_n \times \left ( \Kmat_n \cdot \vvec_n \right ) \right ] = 0. \label{eq:torque-balance}
\end{eqnarray}
On substituting from equations (\ref{eq:vs}) and (\ref{eq:omegas}) into the force and torque 
balance equations, and after some tensorial manipulations to separate $\vvec_{\rm c}$ and 
$\boldsymbol{\Omega}$, we obtain
\begin{eqnarray}
\left [
\begin{array}{cc}
\Cmat_{11} & \Cmat_{12}  \\
\Cmat_{21} & \Cmat_{22} 
\end{array}
\right ] \cdot \left (    
\begin{array}{c}
\vvec_{\rm c}  \\
\boldsymbol{\Omega}
\end{array}
\right ) =
\left (    
\begin{array}{c}
\fvec_{1}  \\
\fvec_{2}
\end{array}
\right ),
\label{eq:determining-matrix-equation}
\end{eqnarray}
which is a more convenient form for motion analysis and control. The elements of the resistance 
matrices $\Cmat_{\alpha\beta}$ and the components of the forcing vectors $\fvec_{\alpha}$ 
($\alpha,\beta=1,2$) have been given in Appendix \ref{app:A}. 
Equations (\ref{eq:determining-matrix-equation}) constitute a system of nonlinear ordinary 
differential equations for the roll-pitch-yaw angles with parametric and external excitations 
through the matrices $\Kmat_n$ and vectors $\fvec_\alpha$. Once $\boldsymbol{\alpha}(t)$ 
is computed in terms of quaternions (see Appendix \ref{app:B}), the position of the 
swimmer is found by integrating $\dot \Xvec_{\rm c}=\vvec_c$.

Quadroar's movements are initialized by setting $s=0$ for the body link and putting 
the disks in appropriate park configurations. This is done in a way that the net drag-induced 
force and torque remain zero while adjusting the disks. Let us define the input control 
signal of the variable $u$ by a simple pulse function in its phase space: $\dot u = \nu_u(t)$ 
for $u_{\rm i} \le u \le u_{\rm f}$ and $\dot u=0$ otherwise. Here $u_{\rm i}$ and $u_{\rm f}$ 
are the initial and terminal conditions of $u$, and $\nu_u$ is a linear or angular velocity. 
We simply define $\dot \vartheta_n = \nu_n(t)$ for disks. One may collect the three variables 
$u_{\rm i}$, $\nu_u$ and $u_{\rm f}$ of each actuator in a single vector and define the 
control signals
\begin{eqnarray}
{\cal S}=\left ( \begin{array}{c}
s_{\rm i} \\ 
\nu_s \\ 
s_{\rm f}
\end{array} \right ), ~~
{\cal D} &=& \left [
\begin{array}{c|c|c|c}
\vartheta_{1{\rm i}} & \vartheta_{2{\rm i}} & \vartheta_{3{\rm i}} & \vartheta_{4{\rm i}} \\
\nu_{1} & \nu_{2} & \nu_{3} & \nu_{4} \\
\vartheta_{1{\rm f}} & \vartheta_{2{\rm f}} & \vartheta_{3{\rm f}} & \vartheta_{4{\rm f}}
\end{array}
\right ].
\end{eqnarray}
Two signals that differ by the signs of $\nu_u$ and have interchanged their initial 
and final states will be denoted by $\pm$ superscripts. For instance, ${\cal D}^{-}$ is obtained 
from ${\cal D}^{+}$ if the elements in the second row of ${\cal D}^{+}$ flip sign, and the first 
and thirds rows are interchanged. The simplest swimming/rotation modes are obtained when 
spatial displacements and rigid-body rotations are decoupled. Below, we further explore 
such operation modes.  

\begin{figure}
\centerline{\hbox{\includegraphics[width=0.45\textwidth]{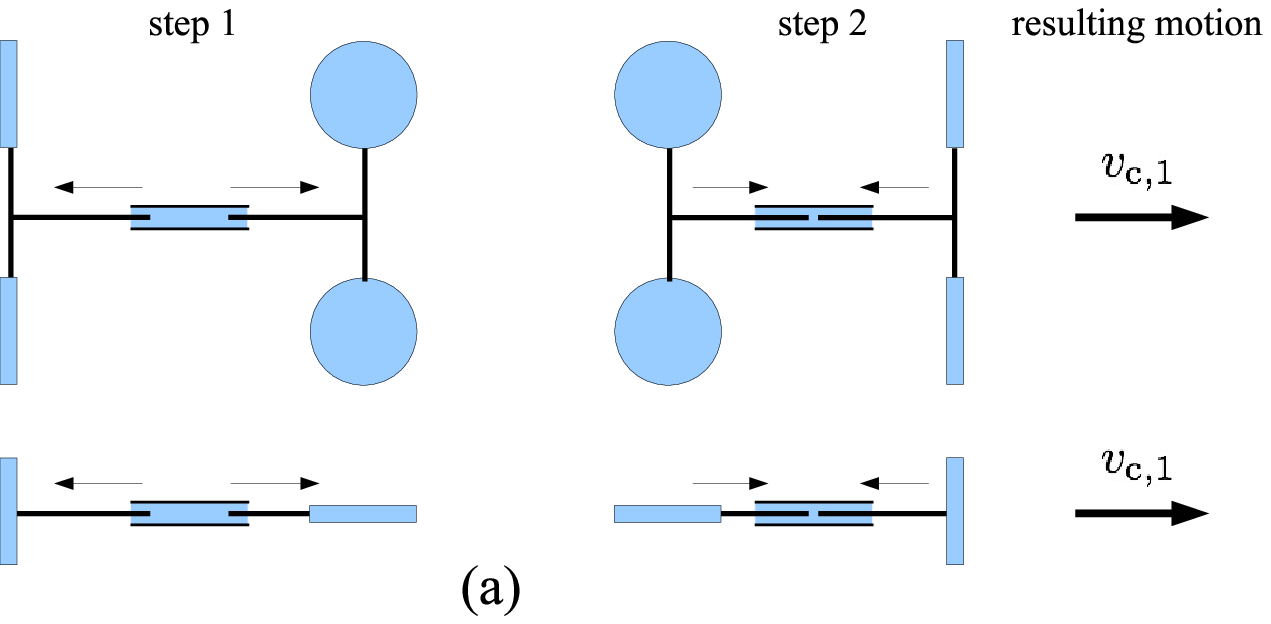} } 
 }
 \vspace{0.1in}
 \centerline{
\hbox{\includegraphics[width=0.45\textwidth]{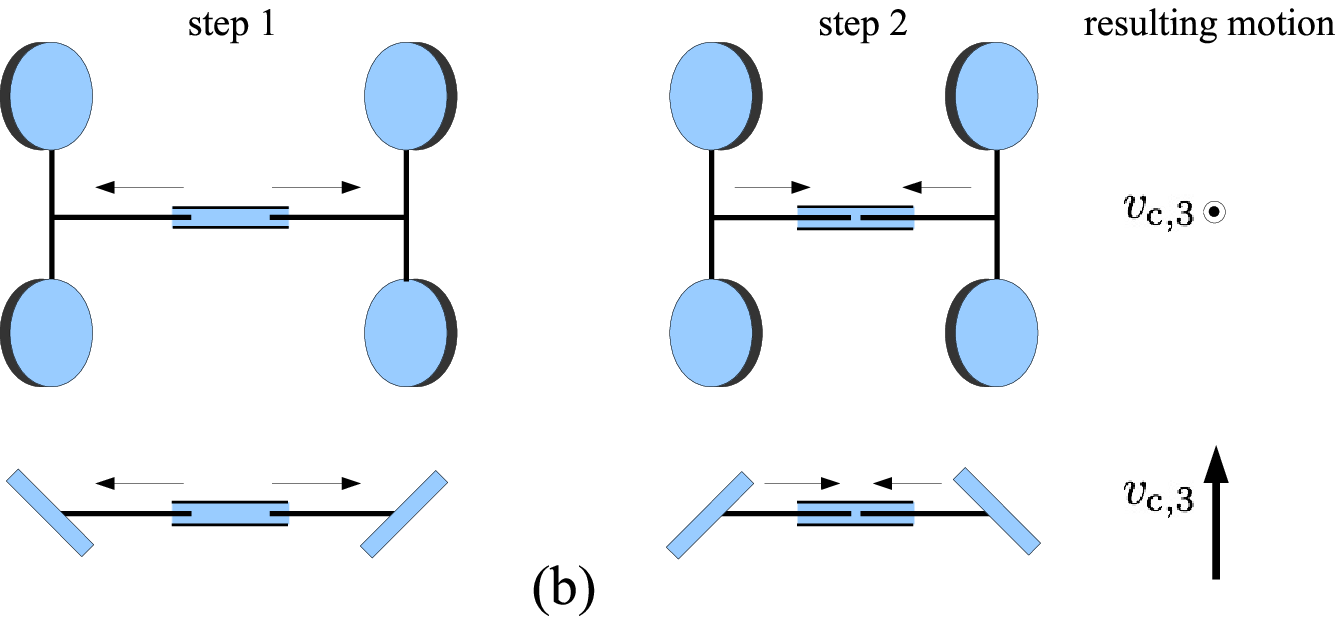} }
 }
\caption{(a) Direct strokes in the $x_1$-direction using a pushme-pullyou type 
sequence. (b) Transverse rigid-body translation in the direction normal to the 
plane of the \texttt{I}-frame. In both swimming modes, disks rotate and acquire 
their desired orientations before the expansion or contraction of the body link begins. 
The simultaneous rotations of the disks before each step are such that the torque 
corresponding to the first (isotropic) term in equation (\ref{eq:torque-balance}) 
vanishes and we get $\boldsymbol{\Omega}=\textbf{0}$.}
\label{fig2}
\end{figure}

The first swimming mode begins with the initializing signal 
\begin{eqnarray}
{\cal D}_0=\left [
\begin{array}{c|c|c|c}
 0 & 0 & 0 & 0 \\
 0 & 0 & \nu_3 & -\nu_3 \\
 0 & 0 & \frac{\pi}{2} & -\frac{\pi}{2}
\end{array}
\right ],
\end{eqnarray}
that puts D3 and D4 in a vertical position (with respect to the \texttt{I}-frame
as shown in the step 1 of Fig. \ref{fig2}(a)) while keeping $\vvec_{\rm c}$ 
and $\boldsymbol{\Omega}$ equal to zero. We then lock the disks and command 
the body link to expand with $\nu_s>0$, which sets ${\cal S}^{+}$. This keeps 
$\boldsymbol{\Omega}={\bf 0}$ and generates a stroke in the $x_1$-direction. 
To compute the advancing velocity $v_{{\rm c},1}$, we write the force balance 
equation in the form
\begin{eqnarray}
&{}& \left ( \Kmat_1+\Kmat_2 \right ) \cdot \left ( \vvec_{\rm c} +\nu_s \evec_1\right ) \nonumber \\
&{}& \qquad + \left ( \Kmat_3+\Kmat_4 \right ) \cdot \left ( \vvec_{\rm c} -\nu_s \evec_1\right )=0,
\label{eq:x-swim}
\end{eqnarray}
where
\begin{eqnarray}
&{}& \Kmat_1 = \Kmat_2 = 
\frac 83 a \cdot {\rm diag} \left [5-1,4,5+1\right ],  \\
&{}& \Kmat_3 = \Kmat_4 =\frac 83 a \cdot {\rm diag} \left [5+1,4,5-1\right ].
\end{eqnarray}
Equation (\ref{eq:x-swim}) then gives $v_{{\rm c},1}=\nu_s/5$, $v_{{\rm c},2}=0$ and $v_{{\rm c},3}=0$. 
Stopping the expansion of the body link at $s=s_{\rm f}$, and successively applying the signals 
\begin{eqnarray}
{\cal D}_1^{+}=\left [
\begin{array}{c|c|c|c}
 0 & 0 & \frac{\pi}{2} & \frac{\pi}{2} \\
 \nu_1 & \nu_1 & -\nu_1 & -\nu_1 \\
\frac{\pi}{2} & \frac{\pi}{2} & 0 & 0
\end{array}
\right ],
\end{eqnarray}
and ${\cal S}^{-}$ (contracting the body link) will generate another step of rectilinear stroke 
with $v_{{\rm c},1}=\nu_s/5$ and without triggering rigid-body tumblings. A full sequence of direct 
strokes has been demonstrated in Fig. \ref{fig2}(a). It is reminiscent of the pushme-pullyou 
sequence, but instead of bladders, utilizes more practical/feasible rotating circular disks. 
Swimming along the $x_1$-axis continues by repeating the 4-tuple sequence:
\begin{eqnarray}
 {\cal S}^{+} \rightarrow {\cal D}_1^{+} \rightarrow {\cal S}^{-} \rightarrow {\cal D}_1^{-}.
\end{eqnarray}
The switchings ${\cal S}^{+}\rightleftarrows {\cal S}^{-}$ and ${\cal D}_1^{+}\rightleftarrows {\cal D}_1^{-}$ 
describe delayed reciprocating actuations: there is a time delay between the termination of the 
body link's expansion and the beginning of its contraction. During this delay time, the disks are 
rotating to reach their target inclinations. Similarly, the disks remain locked when the body link is 
actuated. 

\begin{figure}
\centerline{\hbox{\includegraphics[width=0.45\textwidth]{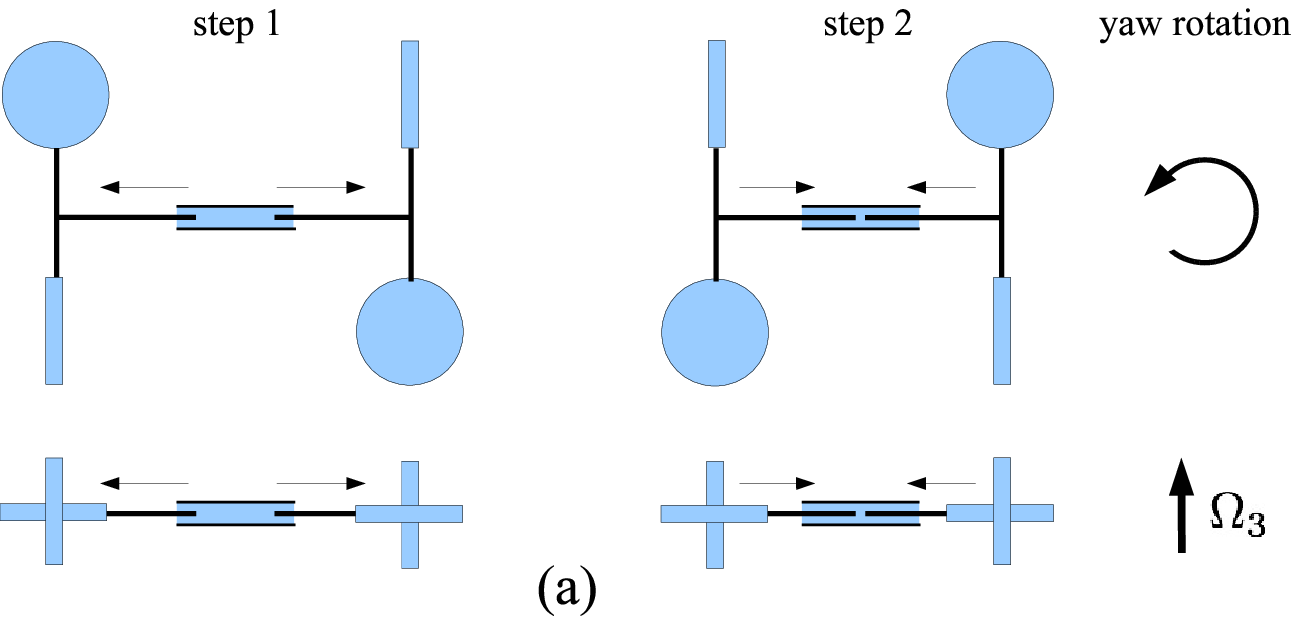} } }\vspace{0.1in}
\centerline{\hbox{\includegraphics[width=0.45\textwidth]{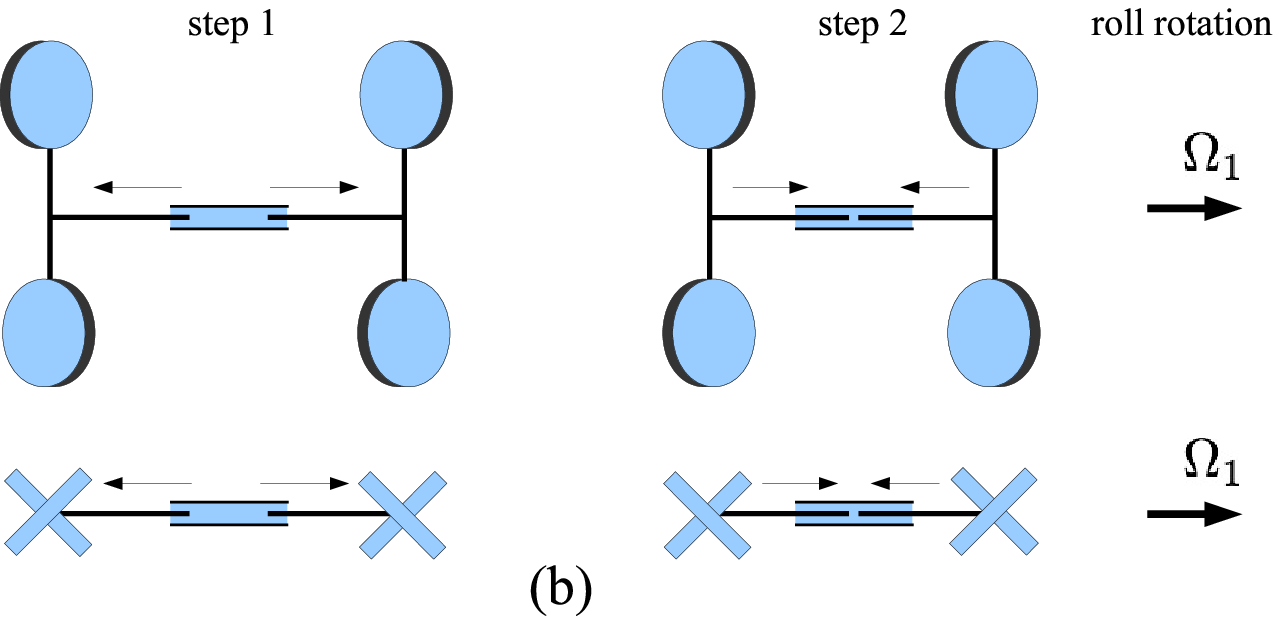} } }
\caption{Steps leading to simple yaw and roll rotations. The vectors of the resulting 
angular velocities have been indicated by arrows on the right hand side of step 2. 
The initial alignments of the disks to their configurations shown in step 1 can be done 
without generating unnecessary tumblings.}
\label{fig3}
\end{figure}

The second swimming mode is a transverse translation normal to the plane of 
the \texttt{I}-frame as shown in Fig. \ref{fig1}(b). The sequence starts with 
initializing the configuration of four disks through the signal 
\begin{eqnarray}
{\cal D}_0=\left [
\begin{array}{c|c|c|c}
 0 & 0 & 0 & 0 \\
 -\nu_1 & -\nu_1 & \nu_1 & \nu_1 \\
 -\vartheta_0 & -\vartheta_0 & +\vartheta_0 & +\vartheta_0
\end{array}
\right ],
\label{eq:initial-signal-vc3}
\end{eqnarray}
and continues with the expansion of the body link by applying ${\cal S}^{+}$. 
This generates a lift with the speed 
\begin{eqnarray}
v_{{\rm c},3}= \frac{\sin(2\vartheta_0)}{5+\cos(2\vartheta_0)} \nu_{s},
\label{e:vc3}
\end{eqnarray}
in the $x_3$-direction. The traveling speed takes the maximum value 
$v_{{\rm c},3}=\nu_s\sqrt{6}/12 \approx \nu_s/5$ when $\cos(2\vartheta_0)=-1/5$. 
The second step of the transverse stroke immediately follows step 1 by 
terminating the expansion of the body link, actuating the disks through 
the control 
\begin{eqnarray}
{\cal D}_3^{+}=\left [
\begin{array}{c|c|c|c}
 -\vartheta_0 & -\vartheta_0 & +\vartheta_0 & +\vartheta_0 \\
 \nu_1 & \nu_1 & -\nu_1 & -\nu_1 \\
 +\vartheta_0 & +\vartheta_0 & -\vartheta_0 & -\vartheta_0
\end{array}
\right ],
\label{eq:signal-vc3}
\end{eqnarray}
then contracting the body link. The resulting speed will be identical to equation (\ref{e:vc3}). 
After the initialization step, transverse strokes persist by repeating the sequence 
\begin{eqnarray}
 {\cal S}^{+} \rightarrow {\cal D}_3^{+} \rightarrow {\cal S}^{-} \rightarrow {\cal D}_3^{-}.
\end{eqnarray}
No further translational modes of swimming decoupled from rotational dynamics exist, 
and the Quadroar cannot perform lateral strokes in the $x_2$-direction because the second 
component of the vector $\Kmat_n \cdot \ell_n \evec_1$ is null for all $n$. Nonetheless, it can 
undergo yaw, pitch and roll rotations, which when followed by direct and transverse strokes
give access to entire three dimensional space. Figure \ref{fig3} demonstrates the steps 
of performing yaw and roll rotations about the $x_3$ and $x_1$ axes, respectively. 
After initialization controls that configure disks to the orientations shown in step 1, 
decoupled rotations are obtained from the 4-tuple sequences
\begin{eqnarray}
 {\cal S}^{+} \rightarrow {\cal D}_{\gamma}^{+} \rightarrow {\cal S}^{-} \rightarrow {\cal D}_{\gamma}^{-},
 ~~ \gamma\equiv {\rm yaw},{\rm roll},
\end{eqnarray}
where the signals ${\cal D}_{\gamma}^{+}$ have been given in Appendix \ref{app:A}, 
and corresponding angular velocities become
\begin{eqnarray}
{\rm Yaw}: ~ \Omega_3 &=& \frac{b}{5b^2+4a^2+4(l+s)^2} \nu_s, \label{eq:yaw} \\
{\rm Roll}: ~ \Omega_1 &=& 
\frac{b \sin\left ( 2\vartheta_0 \right )}{4a^2+b^2[5+\cos(2\vartheta_0)]} \nu_s, \\
s(t) &=& s_{\rm i} + \int_0^{t} \nu_s(\eta) \, d\eta.
\end{eqnarray}
The maximum value of $\Omega_1$ corresponds to $\cos(2\vartheta_0)=-2b^2/(8a^2+10b^2)$
and equals
\begin{eqnarray}
\Omega_{1,{\rm max}} &=& 
\frac{b \sqrt{4a^4+10 a^2 b^2+6 b^4}}{8 a^4+10 a^2 b^2 +12 b^4} \nu_s, \\
&=& \frac{\sqrt{6}}{12} \frac{\nu_s}{b} +{\cal O}(a^2/b^2). \label{eq:max-roll}
\end{eqnarray}
From equations (\ref{eq:yaw}) and (\ref{eq:max-roll}) we conclude that with a constant 
value of $\nu_s$, roll maneuvers are performed faster than yaw ones. The pitch rotation 
does not require the actuation of the body link because it can be generated by the isotropic 
term in equation (\ref{eq:torque-balance}) if all disks rotate synchronously with the same 
phase angle. For $\nu_s = 0$, we find 
\begin{eqnarray}
\Omega_2 &=& -\frac{4 a^2 \nu_1}{4a^2+(l+s)^2 [ 5 + \cos(2 \vartheta_1)]}, 
\label{eq:Omega2-pitch} \\
\vartheta_1&=& \vartheta_2 = \vartheta_3 = \vartheta_4 = \int_{0}^{t} \nu_1(\eta) \, d\eta.
\end{eqnarray}

Motion planning based on the precise control algorithms presented in this section 
can be readily implemented on swimmers in super-millimeter scales, where servo, 
piezo, or IPMC (ionic polymer metal composites) actuators and their controllers are 
available. In sub-millimeter scales, however, the body link and disks may be required 
to continuously driven by light or electromagnetic sources, and the only feasible input 
control is the frequency tuning/modulation of their reciprocating and rotational motions.
Section \ref{sec:frequency-tuning} explores full nonlinear, coupled roto-translatory 
dynamics of the Quadroar without actuation delays.

\section{Continuous actuation}
\label{sec:frequency-tuning}

We excite the body link by the harmonic signal $s=\frac 12 s_{0}[1-\cos(\omega_s t)]$ 
where the maximum expansion $s_{0}$ is one of our control parameters, and $\omega_s$ 
is a constant frequency. We are interested in operation modes when frequency changes 
in the rotational velocities of the disks, and variations in the range of the expansion/contraction 
of the body link, can initiate maneuvers and directed motions near a stationary and 
non-rotating base state defined by $\omega_s=0$ and $\nu_1=\nu_2=-\nu_3=-\nu_4$. 
This study concerns disk rotation rates $\nu_n=(1/k)\omega_s+\Delta \nu_n$ where $k$ is 
an integer number, $\Delta \nu_n$ is a small frequency shift, and disks are initially in phase 
so that $\vartheta_n(0)=0$ ($n=1,2,3,4$). We refer to motions bifurcating from the 
explained base states as retrograde propulsion modes.

\begin{figure}
\centerline{\hbox{\includegraphics[width=0.45\textwidth]{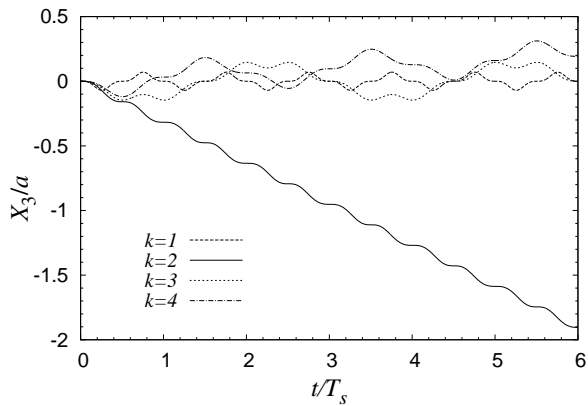} } }
\caption{Distance traveled by the swimmer in the transverse direction for a model 
with $s_0=l/5$ and $b/a=l/a=5$. Dashed, solid, dotted and dash-dotted lines correspond 
to $k=1$, 2, 3, 4, respectively. Initial conditions for the position vector and orientation angles 
are $(X_1,X_2,X_3)=(0,0,0)$ and $(\phi,\theta,\psi)=(0,0,0)$. The time variable is 
normalized to the reciprocating period of the body link $T_s=2\pi/\omega_s$.}
\label{fig4}
\end{figure}
\begin{figure*}
\centerline{\hbox{\includegraphics[width=0.32\textwidth]{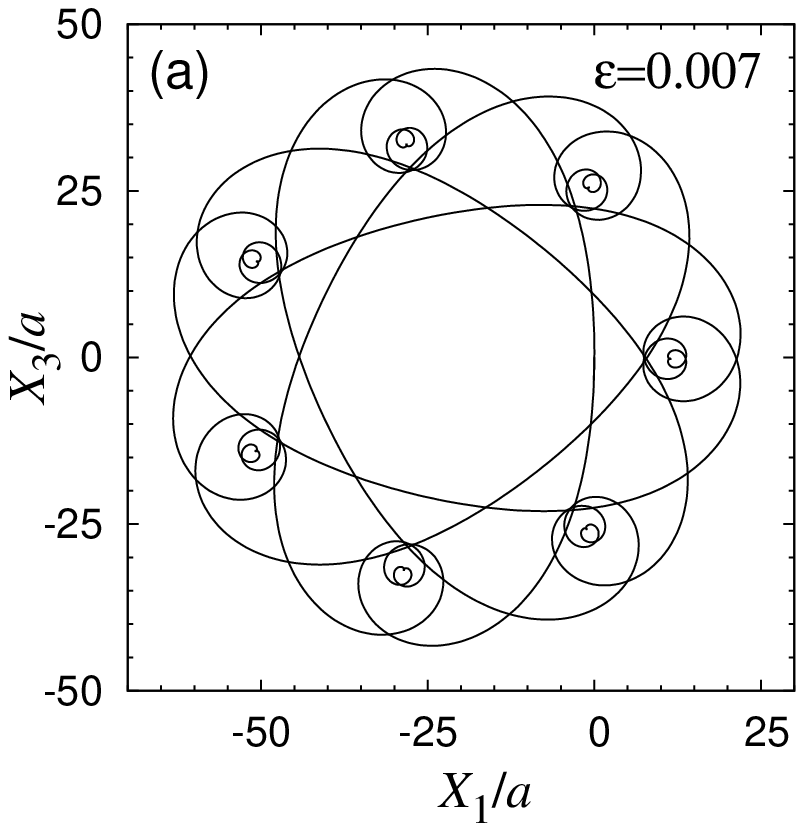} } 
\hbox{\includegraphics[width=0.32\textwidth]{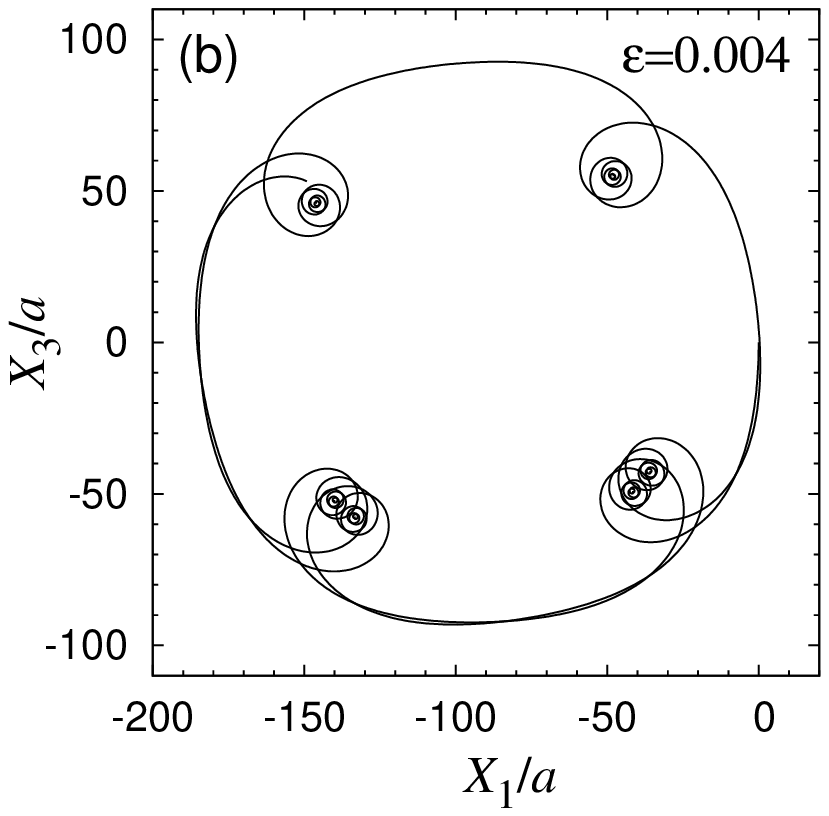} } 
\hbox{\includegraphics[width=0.32\textwidth]{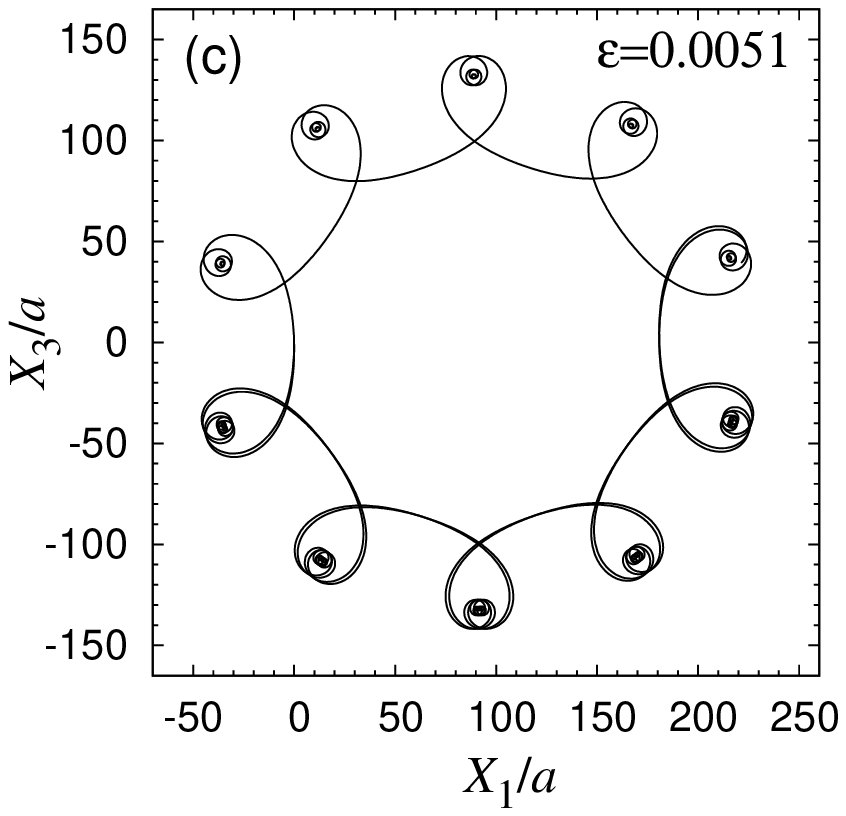} } 
}
\caption{Trajectories of the swimmer in the retrograde propulsion mode. Geometrical 
parameters have been set to $b/a=l/a=5$, the maximum stroke of the body link is 
$s_0=2l/3$, and the rotational velocities of the disks are $\nu_1=\nu_2=\omega_s/2$ 
and $\nu_3=\nu_4=-\omega_s/(2+\epsilon)$. Initial conditions for the position vector 
and orientation angles are $(X_1,X_2,X_3)=(0,0,0)$ and $(\phi,\theta,\psi)=(0,0,0)$, 
respectively. Note the different scales of the panels. The maximum integration times
($t_{\rm max}$) of orbits are also different. (a) An almost periodic rosette orbit 
with $\epsilon=0.007$ and $t_{\rm max}=2010 T_s$ where $T_s=2\pi/\omega_s$. 
(b) A quasi-periodic orbit with 4 multiloop bundles and $\epsilon=0.004$
and $t_{\rm max}=3100 T_s$. (c) A slowly precessing quasi-periodic 
orbit with $\epsilon=0.0051$ and $t_{\rm max}=6000 T_s$.}
\label{fig5}
\end{figure*}

The swimmer undergoes a net directional motion along the body $x_3$-axis if
\begin{eqnarray}
\nu_1=\nu_2=-\nu_3=-\nu_4 =\frac{1}{k} \omega_s,~~ k=2,4,8,\ldots,
\end{eqnarray}
and the traveled distance increases proportional to $s_{0}$. The motion is strictly 
unidirectional only for $k=2$. The center of mass of the swimmer will periodically 
move in the transverse direction with a displacement amplitude $\sim {\cal O}(a)$ 
if $k$ be an odd integer. This simple rectilinear mode of swimming with continuous 
actuation is a time-varying version of the sequence shown in Fig. \ref{fig2}(b).
For initially in phase disk rotations, substituting from $\vartheta_0=-\omega_s t$
and $\nu_s=\dot s=s_0 \, \omega_s \sin(\omega_s t)$ into equation (\ref{e:vc3}) 
and integrating over $t$ gives 
\begin{eqnarray}
X_3(t)=-s_0 \, \omega_s \int_{0}^{t} \frac{\sin(\omega_s \eta)\sin(2\omega_s \eta/k)}{5+\cos(2\omega_s \eta/k)} \, d\eta.
\end{eqnarray}
Figure \ref{fig4} shows the variation of $X_3(t)$ for a model with $s_0=l/5$, $b/a=l/a=5$ 
and for several values of $k$. The body $x_3$-axis is aligned with the inertial $X_3$-axis 
due to our special choice of the swimmer's initial orientation. For $k=2$ one can show that
\begin{eqnarray}
X_3(t) &=& \frac{5 s_0 \, \omega_s t}{2} 
-{\frac {s_{{0}}\tan \left( \omega_s t/2 \right) }{ 1+\left [\tan \left(
\omega_s t /2 \right)  \right ] ^{2}}} \nonumber \\
&{}& -2\, s_{{0}}\sqrt {6}\arctan \left( \sqrt{6}/3\,\tan
 \left( \omega_s t/2 \right) \right).
\end{eqnarray}
Sampling $X_3(t)$ at the discrete times $t=m\pi/\omega_s$ ($m=1,2,3,\ldots$) yields the recursive relation 
\begin{eqnarray}
X_3(m) &=& X_3(m-1) + \left ( \frac{5}{2}-\sqrt{6} \right )\pi \, s_0, \\
X_3(m) & \equiv & X_3 (t=m\pi/\omega_s).
\end{eqnarray}
Therefore, the longest step that the Quadroar can take in the retrograde propulsive mode 
over a cycle of the body link's actuation is $(5-2\sqrt{6})\pi s_0 \approx 0.317 s_0$.

\begin{figure*}
\centerline{\hbox{\includegraphics[width=0.32\textwidth]{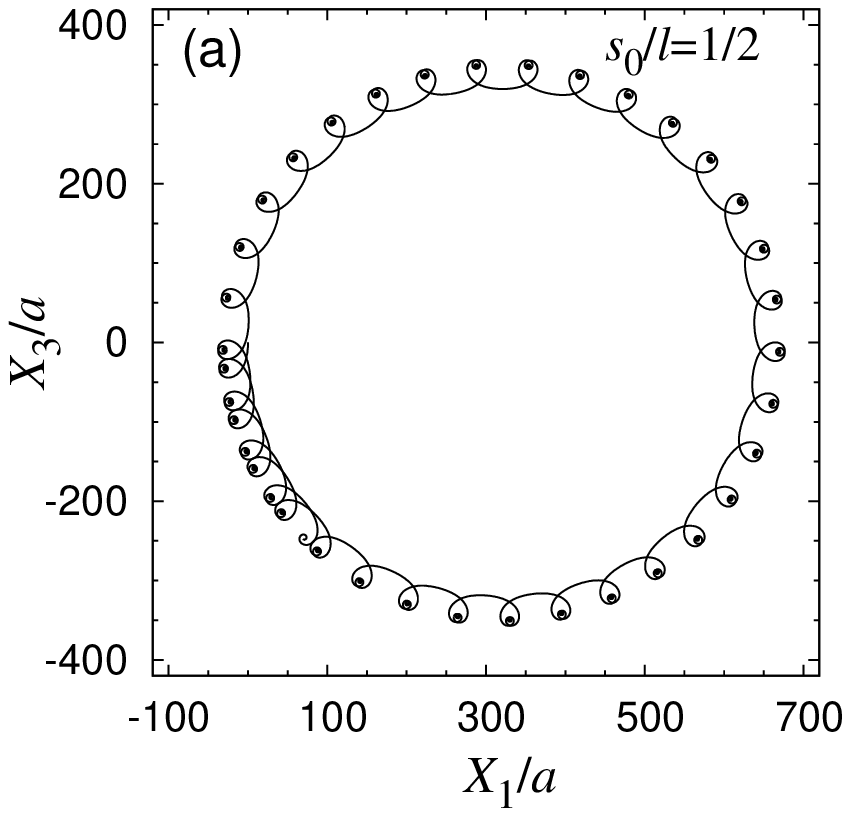} } 
\hbox{\includegraphics[width=0.32\textwidth]{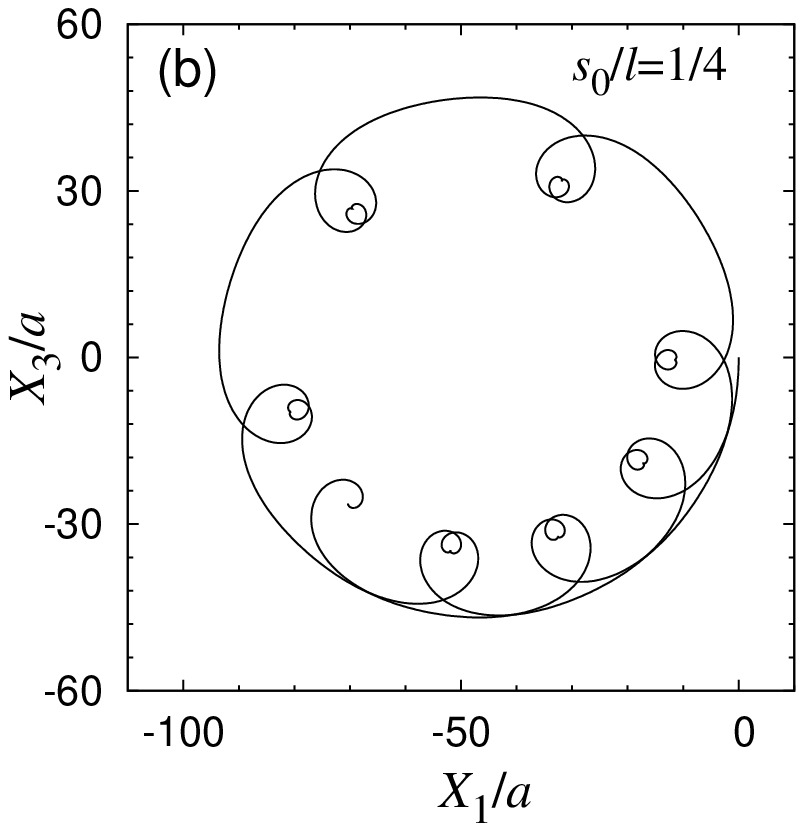} } 
\hbox{\includegraphics[width=0.32\textwidth]{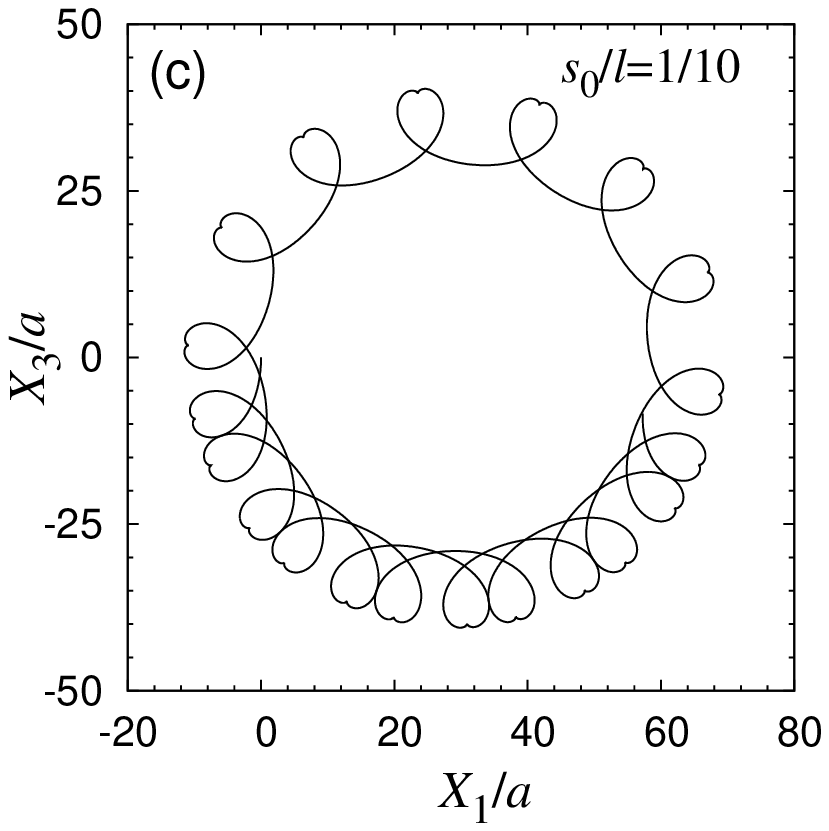} } 
}
\caption{Same as Fig. \ref{fig5} but for models with $\epsilon=0.005$ and three 
choices of $s_0/l$. All orbits are quasi-periodic. (a) $s_0/l=1/2$ and $t_{\rm max}=15000 T_s$. 
(b) $s_0/l=1/4$ and $t_{\rm max}=3000 T_s$. (c) $s_0/l=1/10$ and $t_{\rm max}=8000 T_s$.}
\label{fig6}
\end{figure*}

Interesting motions occur when the rotational speeds of the disks are detuned, and the 
swimmer tumbles as it moves in the three dimensional space. The simplest roto-translatory 
mode of swimming is when the motion is confined to a plane, and only the pitch angle 
$\theta$ varies over time. Figure \ref{fig5} shows the Quadroar's trajectory for $b/a=l/a=5$, 
$s_0= 2l/3$, and rotational frequencies $\nu_1=\nu_2=\omega_s/2$ and 
$\nu_3=\nu_4=-\omega_s/(2+\epsilon)$. The orbit corresponding to $\epsilon=0.007$ 
is an almost periodic rosette with 7 lobes and triple hierarchical loops at the turning points
(Fig. \ref{fig5}(a)). The innermost loop is a cardioid. By decreasing $\epsilon$ to $0.004$ 
we obtain a quasi-periodic orbit containing 4 centrophillic multiloops on its periphery
(Fig. \ref{fig5}(b)). Each multiloop bundle now consists of 6 loops. The motion on quasi-periodic 
orbits that we find exhibit two generic phases: lingering while the swimmer undergoes a 
cardioid or multiloop turn, followed by a propulsive phase of long-distance travel. 
The mean radii of orbits, the number of loops in lingering phases and the centrophillic 
or centrophobic alignments of the multiloop bundles nonlinearly depend on $s_0$ 
and $\epsilon$. For instance, setting $\epsilon=0.0051$ yields a slowly precessing 
orbit with 10 centrophobic multiloop bundles (Fig. \ref{fig5}(c)). 

\begin{figure}
\centerline{\hbox{\includegraphics[width=0.22\textwidth]{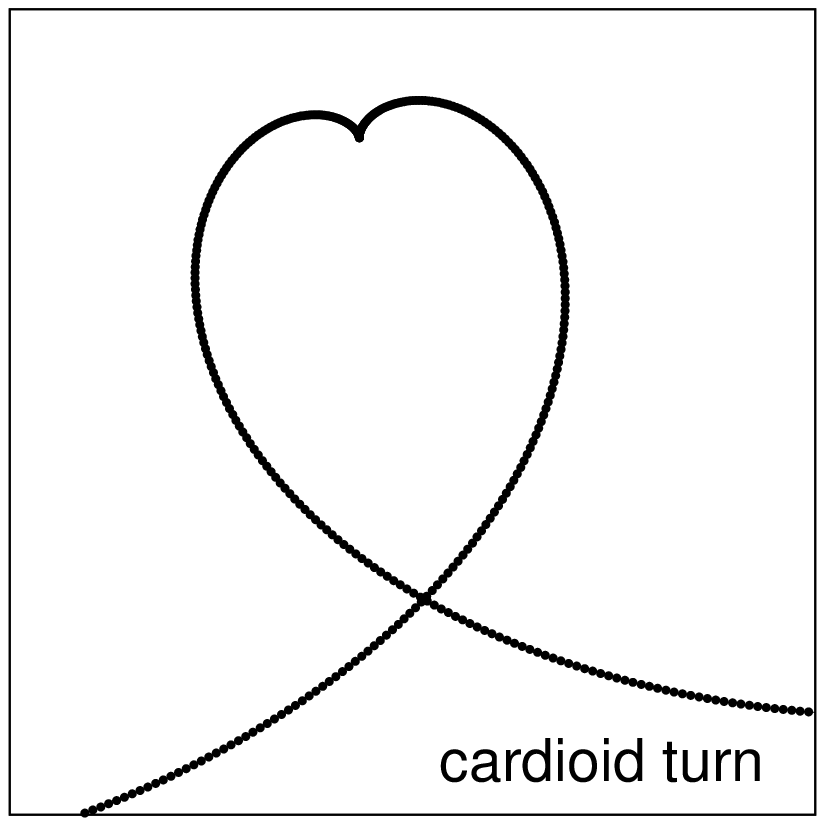} } 
\hbox{\includegraphics[width=0.22\textwidth]{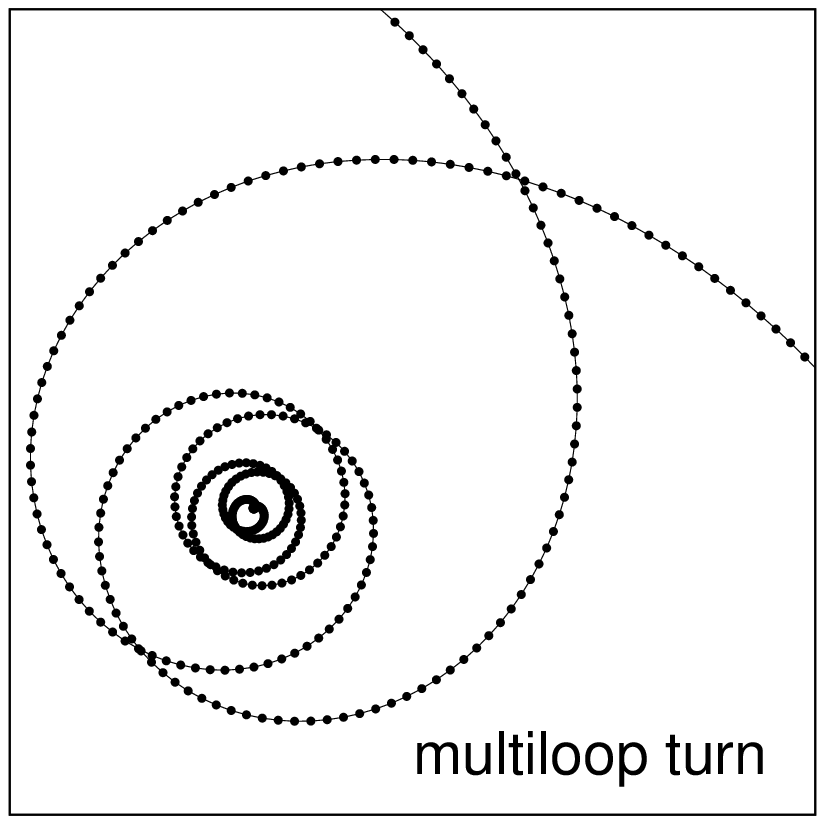} } 
}
\caption{Cardioid (left) and multiloop (right) turning paths. In both cases the 
trajectory becomes cuspy and non-differentiable at one position. The innermost 
loop of the multiloop bundle is a cardioid.}
\label{fig7}
\end{figure}

We conduct further survey in the parameter space by keeping $\epsilon$ fixed and 
varying $s_0$. Figure \ref{fig6} shows that the mean diameter of the quasi-periodic 
trajectory is boosted significantly by taking $s_0=l/2$ and $\epsilon=0.005$, but it
drops by decreasing $s_0$. The pattern demonstrated in Fig. \ref{fig6}(c) has the 
least number of loops at the periphery of the orbit. Its turning loop is a single cardioid. 
Figure \ref{fig7} displays a single cardioid and a multiloop structure. The existence of 
single or multiple loop bundles together with long-distance swims can be explained 
in the context of resonances. Since the front and rear disks rotate with different 
angular velocities, even though the frequency differences are small, phase shifts  
occur over time. The Quadroar propels on an almost straight line when the phase 
angles are such that
\begin{eqnarray}
\vartheta_1 \approx -\vartheta_3,~~\cos(2\vartheta_1) \approx -1/5,
\end{eqnarray}
and the disks are in the relative configurations of Fig. \ref{fig2}(b) needed for the 
transverse propulsion mode. It is remarked that $\vartheta_2=\vartheta_1$ and
$\vartheta_4=\vartheta_3$ for the planar swimming modes of Figs \ref{fig5} and 
\ref{fig6}. The swimmer moves on a multiloop pattern when all $\vartheta_n$ 
have the same sign. Motions on curved paths will thus be intermediary 
phases with ${\rm sign}(\vartheta_1) =-{\rm sign}(\vartheta_3)$. 

\begin{figure}
\centerline{\hbox{\includegraphics[width=0.46\textwidth]{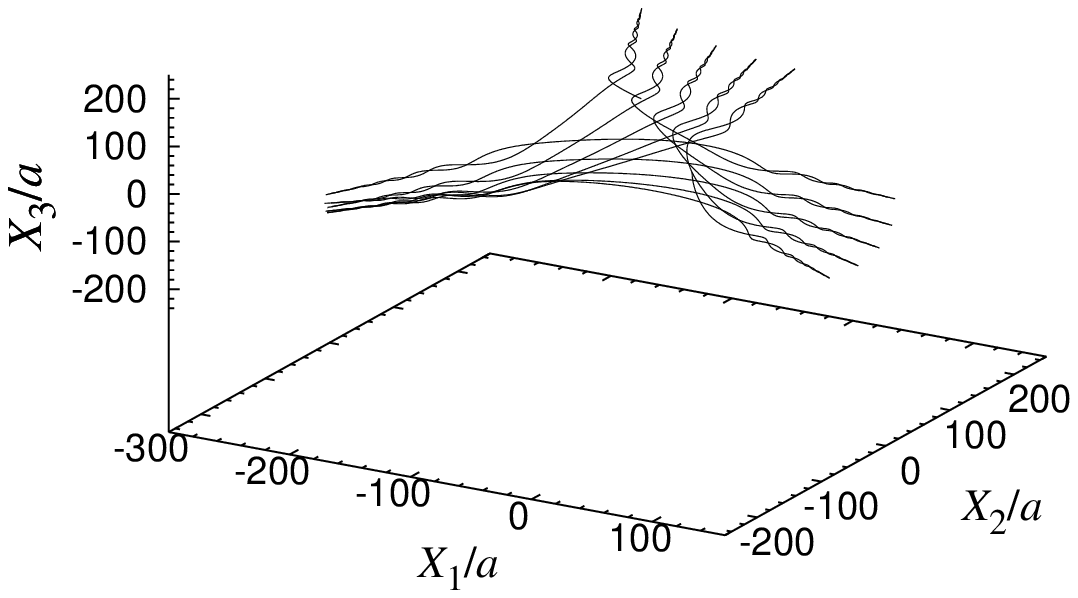} } }
\hspace{0.1in}
\centerline{\hbox{\includegraphics[width=0.47\textwidth]{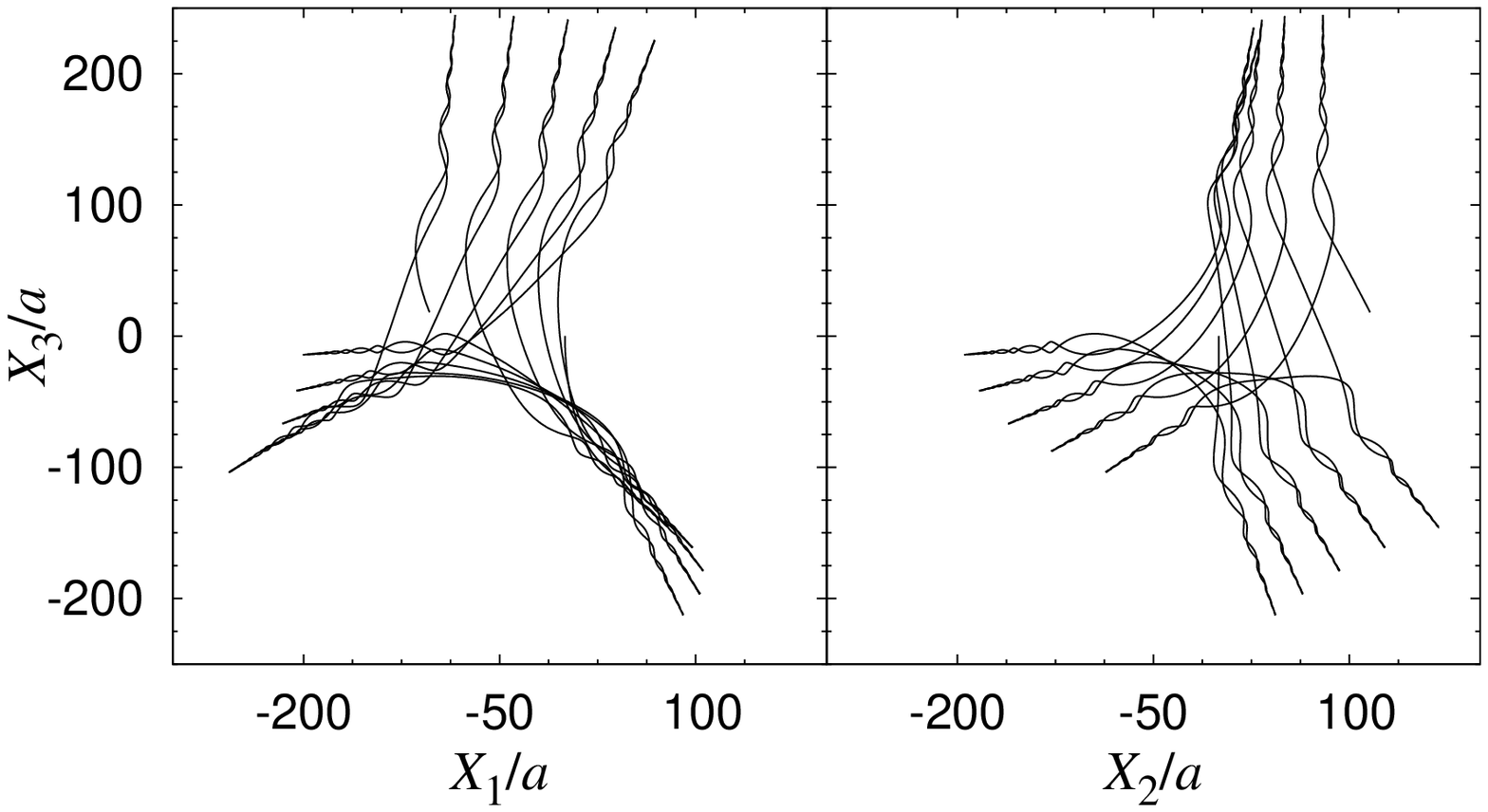} } 
}
\caption{A three dimensional quasi-periodic orbit and its projections on the 
$(X_1,X_3)$ and $(X_2,X_3)$ coordinate planes for a model with $b/a=l/a=5$, 
$s_0=l/2$, $\nu_1=\omega_s/2$, $\nu_2=\omega_s/(2+\epsilon)$, $\nu_3=-\omega_s/2$
and $\nu_4=-\omega_s/(2-\epsilon)$ where $\epsilon=0.002$. The integration time is
$t_{\rm max}=15000 T_s$ and the swimmer starts its motion from $(X_1,X_2,X_3)=(0,0,0)$. 
This is a swimming mode with retrograde propulsion and has screw-shape lobes with 
cusped turning points. The projections of the lobes precess in the coordinate planes.}
\label{fig8}
\end{figure}

Our numerical simulations show that the rotational speeds $\nu_1=-\nu_3=\omega_s/k$ 
and $\nu_2=-\nu_4=\omega_s/(k+\epsilon)$ induce generally quasi-periodic motions 
in the frontal $(x_2,x_3)$-plane of the swimmer. Similar to motions in the $(x_1,x_3)$-plane,
the mean radii of orbits can be orders of magnitudes larger than the size of the swimmer. 
A full three dimensional motion occurs when no mirror and central symmetries exist
in the magnitudes of $\nu_n$, i.e., $\nu_1\not =\nu_2$ and 
$\vert \nu_1\vert \not =\vert \nu_3 \vert$. This extends the number of control 
parameters to 5, and we need to tune four frequency ratios/shifts along with $s_0$ 
to generate desired bounded orbits. The most interesting combination of disk
rotation rates is $\nu_1=\omega_s/2$, $\nu_2=\omega_s/(2+\epsilon)$, 
$\nu_3=-\omega_s/2$ and $\nu_4=-\omega_s/(2-\epsilon)$, which together with 
the stroke length $s_0=l/2$ yield the three dimensional quasi-periodic orbit of 
Fig. \ref{fig8}. It is seen that the orbit has three screw-shaped lobes whose 
projections on the $(X_1,X_3)$ and $(X_2,X_3)$ coordinate planes precess. 
The orbit makes an inclined structure in the $(X_1,X_2)$-plane. The screw-shape 
trajectory is due to a combination of roll and yaw rotations and the translation along 
the body $x_3$-axis. In this operation mode the swimmer also performs pitch 
rotations when the phase angles of all disks have the same sign. 
The inclination of the resulting orbit in the inertial frame depends on the initial 
orientation of the swimmer that can be adjusted by simple maneuvers introduced 
in section \ref{actuation modes}. Further intentional shifts in the frequencies of the 
disks can also put the swimmer on transitional paths and change its attitude. 
We note that the spatial orientation of orbits with initially non-zero and unequal 
phase angles are not necessarily identical to the ones with initially zero phases
that we experimented here.


\section{discussions}
\label{sec:discuss}

Swimming in more than one dimensions involves the attitude and angular 
velocity calculations of the swimmer. The set of four disks together with the 
thin and frictionless \texttt{I}-frame result in simple translation and resistance 
matrices, which were used to generate basic displacements and attitude 
maneuvers through sending delayed pulses to the actuators. Since spins 
and translations can be performed independently, we have basically a 
zero-turning-radius swimmer. These two properties offer a wide range of 
capabilities for non-contact fluidic manipulation \cite{Floyd09}. 

Coupled roto-translatory motions with the continuous operations of the actuators 
led to complex yet geometrically well-defined orbital patterns, which were found 
to be sensitive to the stroke length of the body link. Our study is restricted to 
rotational frequencies of disks that satisfy $\nu_n=\omega_s/k+\Delta \nu_n$, 
and for operation modes that require disks to start their rotation in phase. 
Even with these limited sets of model parameters, the Quadroar showed to 
have a rich dynamics. In continuous retrograde propulsion modes it moves on 
periodic and quasi-periodic orbits with mean diameters orders of magnitude 
larger than its body size, and its trajectory becomes dense in the invariant 
measure of quasi-periodic orbits. This exceptional capability makes it a 
candidate for accurate search and scanning purposes in low Reynolds 
number regimes. It is worth noting that quasi-periodic orbits have also 
been observed in the motion of planetary and stellar systems \cite{BT08}. 
Both planar and three dimensional quasi-periodic orbits involved slow 
turning motions on non-differentiable curves, as shown in Fig. \ref{fig7}. 
A full understanding of the correlation between the topology of orbits and 
the control parameters $s_0$ and $\nu_n/\omega_s$, and the existence 
of chaotic movements are interesting problems for future work.

Swimmers induce a flow field that contributes to the mixing of the surrounding fluid. 
Such a mixing in low Reynolds number regimes benefits from the long-range 
nature of the velocity field created by the swimmer, but suffers from the lack of 
turbulence. Mixing in low Reynolds number is important as it enhances the rate 
of the particle diffusion in the fluid which results in, for example, a more efficient 
nutrient supply to microorganisms \cite{Pushkin2013}. It is, in fact, believed 
that the rational behind the motion of microorganisms, besides immediate 
survival, is partly to optimally distribute the nutrients \cite{Gregoire2001}. 

Low Reynolds number swimmers are known to enhance mixing. Over the 
surface of a soap film it has been observed that bacteria induce super-diffusion 
to the nearby fluid particles \cite{Gregoire2001}. Motile bacteria also observed 
to enhance the mixing at the interface of two streams of fluid \cite{Kim2004,Kim2007a}, 
and a flat rigid surface covered with restrained bacteria can significantly increase 
the mixing of adjacent fluid particle \cite{Darnton2004a}. While the details of 
flow pattern produced by the Quadroar is beyond the scope of this article and 
deserves an independent investigation, we would like to comment that the 
broad family of rich sweeping patterns, that can be periodic, quasi-periodic 
and potentially chaotic, can result in a wide range of complex flow fields 
capable of driving an efficient mixing, which is currently achieved by boundary 
actuation \cite{Mathew07} or externally actuated stirring \cite{Couchman10}. 
Rotating disks of the Quadroar can be considered as micro-impellers \cite{Yang2013,Hessel2005} 
that not only directly stir the flow, but also propel the Quadroar forward along 
specific yet complex trajectories to achieve maximum efficiency stirring. If stirring is 
the only objective, disks may be further modified but this will clearly also affect 
the overall trajectory of the Quadroar, and therefore a careful optimization 
must be carried out beforehand.

With current developments in nano sciences and technology, the Quadroar 
seems to be a realizable swimmer in molecular scales. For more than a decade, 
several groups of organic chemists have been designing and synthesizing 
molecular motors that work by electrochemical, photochemical or thermal 
actuations \cite{K05,MS09}. One of the prominent examples in this line, 
is the unidirectional light-driven motor of \citet{K99} based on symmetric 
biphenanthrylidenes, and its second generation \cite{K02} that can be 
connected to a supporting frame. The unidirectional motion of this nano 
motor is maintained by a stereogenic center, and the helix inversion of the 
molecule through irradiation or heating. These motors can be used as the 
driving units of various complexes, including nano cars with carborane 
wheels \cite{C12}. A recently experimented electrically driven wheeled 
molecule uses four unidirectional rotary motors to move on metal surfaces. 
A similar concept can be used to synthesize a nano-scale Quadroar, but we 
also need a light-driven molecular actuator that functions like the telescopic 
body link of our swimmer. This central unit can be designed in a number of 
ways: (i) using an azobenzene chassis that expands and contracts through 
cis-trans photoisomerization. (ii) adapting a mechano-sensitive channel 
protein that can open and close by successive irradiation of ultraviolet and 
visible lights \cite{Kocer05}. Azobenzene chromophores have been successfully 
applied in synthesizing photoactive nanoworms \cite{ST08} and molecular 
scissors \cite{M03}, and seem to be more feasible units for generating the 
open-close function of the body link of a nano-scale Quadroar.

\appendix
\section{Rotation and Transfer Matrices}
\label{app:A}
The transformation matrices for roll-pitch-yaw rotations are defined as
\begin{eqnarray}
{\rm Yaw:~} \Rmat_{x_3}(\psi) &=& \left [
\begin{array}{ccc}
\cos \psi ~ &  \sin \psi & ~ 0  \\
-\sin \psi ~ & \cos \psi  & ~ 0  \\
0 ~ & 0  & ~ 1  
\end{array}
\right ],
\end{eqnarray}
\begin{eqnarray}
{\rm Pitch:~} \Rmat_{x_2}(\theta) &=& \left [
\begin{array}{ccc}
\cos \theta ~ & 0 & ~ -\sin \theta   \\
0 ~ & 1  & ~ 0  \\
\sin \theta ~ & 0  & ~ \cos \theta
\end{array}
\right ], 
\end{eqnarray}
\begin{eqnarray}
{\rm Roll:~} \Rmat_{x_1}(\phi) &=& \left [
\begin{array}{ccc}
1 ~ &  0 & ~ 0  \\
0 ~ & \cos \phi  & ~ \sin \phi  \\
0 ~ & -\sin \phi  & ~ \cos \phi  
\end{array}
\right ].
\end{eqnarray}
The elements of the resistance matrices $\Cmat_{\alpha\beta}=C_{\alpha\beta,ij}\evec_i \evec_j$ 
and the components of the forcing vectors $\fvec_{\alpha}=f_{\alpha,i}\evec_i$ ($\alpha,\beta=1,2$) are:
\begin{eqnarray}
\Cmat_{11} &=& \sum_{n=1}^{4} \Kmat_{n} \cdot \Rmat,
\end{eqnarray}
\begin{eqnarray}
C_{12,ij} &=& \sum_{n=1}^{4} K_{n,ik} r_{n,m} \varepsilon_{mkj},
\end{eqnarray}
\begin{eqnarray}
C_{21,ij} &=& \sum_{n=1}^{4} \varepsilon_{imk} r_{n,m} K_{n,kl} R_{lj},
\end{eqnarray}
\begin{eqnarray}
C_{22,ij} &=& \sum_{n=1}^{4} \left (  \frac{32 a^3}{3} \delta_{ij} +
\varepsilon_{imk} \varepsilon_{ljp} r_{n,m} r_{n,p} K_{n,kl} \right ), ~~~~
\end{eqnarray}
\begin{eqnarray}
f_{1,i} &=& - \sum_{n=1}^{4} K_{n,i1} \dot \ell_n,
\end{eqnarray}
\begin{eqnarray}
f_{2,i} &=& -\sum_{n=1}^{4} \left ( \frac{32 a^3}{3} \delta_{2i} \dot \vartheta_n
+\dot \ell_n r_{n,j} K_{n,k1} \varepsilon_{ijk} \right ),
\end{eqnarray}
where $\delta_{ij}$ is Kronecker's delta and $\varepsilon_{ijk}$ is the permutation symbol.
With reciprocating actuators, periodic control signals for the yaw and roll rotations are 
computed from 
\begin{eqnarray}
{\cal D}_{\rm yaw}^{+} &=& \left [
\begin{array}{c|c|c|c}
 -\frac{\pi}{2} & 0 & 0 & +\frac{\pi}{2} \\
 \nu_1 & \nu_1 & -\nu_1 & -\nu_1 \\
 0 & \frac{\pi}{2} & -\frac{\pi}{2} & 0
\end{array}
\right ], \label{eq:signal-omega3}
\end{eqnarray}
\begin{eqnarray}
{\cal D}_{\rm roll}^{+} &=& \left [
\begin{array}{c|c|c|c}
 -\vartheta_0 & +\vartheta_0 & +\vartheta_0 & -\vartheta_0 \\
 \nu_1 & -\nu_1 & -\nu_1 & \nu_1 \\
 +\vartheta_0 & -\vartheta_0 & -\vartheta_0 & +\vartheta_0
\end{array}
\right ]. \label{eq:signal-omega1}
\end{eqnarray}

\section{Unit Quaternions}
\label{app:B}
While integrating the equations $\dot{\boldsymbol{\alpha}} = \Tmat^{-1} \cdot \boldsymbol{\Omega}$ 
to determine the orientation of the swimmer, a geometrical singularity occurs at $\theta=\pm \pi /2$. One way of resolving such singularities is to switch between roll-pitch-yaw and other sets of Euler angles. Alternative approach, which we adopt in this study, is to express the attitude dynamics 
in terms of the unit quaternions:
\begin{eqnarray}
\qvec = 
\left (
\begin{array}{c}
q_0    \\
q_1    \\
q_2    \\
q_3   
\end{array}
\right ) =
\left (
\begin{array}{c}
c_{\frac{\phi}{2}} c_{\frac{\theta}{2}} c_{\frac{\psi}{2}} + s_{\frac{\phi}{2}} s_{\frac{\theta}{2}} s_{\frac{\psi}{2}}  \\
-c_{\frac{\phi}{2}} s_{\frac{\theta}{2}} s_{\frac{\psi}{2}} + s_{\frac{\phi}{2}} c_{\frac{\theta}{2}} c_{\frac{\psi}{2}}  \\
c_{\frac{\phi}{2}} s_{\frac{\theta}{2}} c_{\frac{\psi}{2}} + s_{\frac{\phi}{2}} c_{\frac{\theta}{2}} s_{\frac{\psi}{2}}    \\
c_{\frac{\phi}{2}} c_{\frac{\theta}{2}} s_{\frac{\psi}{2}} - s_{\frac{\phi}{2}} s_{\frac{\theta}{2}} c_{\frac{\psi}{2}}   
\end{array}
\right ),
\end{eqnarray}
whose dynamics is governed by
\begin{eqnarray}
\dot \qvec =  \frac 12
\left [
\begin{array}{cccc}
0 & -\Omega_1  & -\Omega_2 & -\Omega_3  \\
\Omega_1 & 0 & \Omega_3 & -\Omega_2   \\
\Omega_2 & -\Omega_3 & 0 & \Omega_1   \\
\Omega_3 & \Omega_2 & -\Omega_1 & 0   
\end{array}
\right ] \cdot \qvec.
\end{eqnarray}
After integrating these equations in the time domain, inverse transformation to the space of roll-pitch-yaw angle is carried out using the following relations
\begin{eqnarray}
\boldsymbol{\alpha} =
\left (
\begin{array}{c}
 {\rm atan2}( 2 q_2 q_3+ 2 q_0 q_1,q_3^2-q_2^2-q_1^2+q_0^2)   \\
 -\arcsin(2 q_1 q_3 -2 q_0 q_2)   \\ 
 {\rm atan2}( 2 q_1 q_2 + 2q_0 q_3,q_1^2 +q_0^2 -q_3^2 - q_2^2) 
\end{array}
\right ). ~~~~
\end{eqnarray}

\end{document}